\documentclass[aps,prl,twocolumn,superscriptaddress,groupedaddress]{revtex4}  
\usepackage{graphicx}  
\usepackage{dcolumn}   
\usepackage{bm}        
\usepackage{amssymb}   

\def\ba{\begin{eqnarray}}
\def\ea{\end{eqnarray}}
\def\beq{\begin{eqnarray}}
\def\eeq{\end{eqnarray}}

\def\S{{\Sigma}}

\def\S{\Sigma}

\def\L*{{\cal L}_*}
\def\L{\mathcal{L}}
\def\({\left(}
\def\){\right)}

\def\nn{\nonumber}

\def\<{\langle}
\def\>{\rangle}
\newcommand{\eqref}[1]{(\ref{#1})}

\begin{document}

\title{Supersymmetric Grand Unification with Light Color-Triplet}
\author{Lasha Berezhiani}
\address{Center for Cosmology and Particle Physics,
Department of Physics, New York University,
NY, 10003, USA}
\date{\today}


\begin{abstract}

We construct a natural model of the supersymmetric $SU(6)$ unification, in which the symmetry breaking, down to the standard model gauge group, results in the number of pseudo-Nambu-Goldstone superfields with interesting properties. Namely, besides the Higgs doublet-antidoublet pair which is responsible for the electroweak phase transition, the Nambu-Goldstone sector consists of multiplets in the anti- and fundamental representations of $SU(5)$. While being strictly massless in the supersymmetric limit, they acquire the weak scale masses as a result of its breaking. The color-triplet components of this light sector could, in principle, mediate an unacceptably fast proton decay; however, because of the natural $\text{TeV}/M_{\text{GUT}}$ suppression of the Yukawa couplings to the light quarks and leptons, their existence is compatible with the experimental bound on proton lifetime. This suppression is made further interesting, since it results  in the lifetime, of the lightest of the above-mentioned colored particles from $1$sec to $1$day, long enough for it to appear stable in the detector. Furthermore, we argue that the accommodation of the color-triplet pseudo-Nambu-Goldstones, without fine-tuning or contradicting observations, implies $SU(6)$ unification.

\end{abstract}

\maketitle



\section*{I. Introduction}
One of the most elegant solutions to the hierarchy problem is supersymmetry \cite{Maiani}. It also predicts the unification of gauge coupling constants at energies $\sim10^{16}GeV$ \cite{crossing}. Nevertheless, the actual embedding of the standard model gauge group into the more powerful one is quite peculiar. Namely, one needs to explain the relative lightness of the Higgs doublets as compared to the GUT scale, while keeping their triplet partners heavy (with mass $\sim M_{GUT}\equiv 10^{16}GeV$). Otherwise, they would mediate an unacceptably fast proton decay \cite{dim5}. This is the infamous doublet-triplet splitting problem. There are several ways to achieve this splitting. One of the most attractive ways is the so called GIFT (goldstone instead of the fine-tuning) mechanism \cite{Inoue}-\cite{DP}, in which the lightness of the MSSM Higgs doublets is achieved by their pseudo-goldstone nature. 

In particular, it was noticed in \cite{Inoue,Anselm} that the Higgs sector of the minimal $SU(5)$ theory, which consists of $24 + 5 + \bar{5}$, upon addition of an extra singlet, can be completed into the content of the adjoint representation of $SU(6)$, the $SU(5)$ decomposition of which is $35 = 24 + 5+ \bar{5}+1$. Moreover, by appropriate adjustment of the parameters in the Higgs superpotential, the latter can be written in a globally $SU(6)$ invariant form, as a function of a single 35-plet. The observation of \cite{Inoue, Anselm} was that, once the local symmetry is broken down to the standard model gauge group, the pattern of the global symmetry breaking is such that it produces two weak-doublet pseudo-Nambu-Goldstones (pNG). However, at the level of this construction there is no rationale for the $SU(6)$ global symmetry and it appears that one fine-tuning is replaced by the other.

The resolution of this puzzle was provided in \cite{BD}. The key point being that such a global symmetry can result from the $SU(6)$ gauge invariant theory, in which two Higgs sectors are essentially decoupled from each other. The minimal Higgs content, necessary for the desired symmetry breaking, consists of an adjoint and a pair of anti- and fundamental representations: $\S \equiv 35$, $H\equiv 6$ and $\bar H \equiv \bar 6$. The theory is arranged in such a way that the superpotential splits in two sectors
\beq
W=W(\S)+W(H,\bar H).
\label{w1}
\eeq
As a result, the global symmetry is effectively enhanced to $SU(6)_\S \times U(6)_H$, corresponding to the independent transformations of $\S$ and $H$'s; it must be emphasized that this is the symmetry of merely the Higgs sector and is explicitly broken by the Yukawa couplings. The theory admits the supersymmetric ground state
\beq
\< \S \>=0, \qquad \< H \>=V_H \text{diag}(1,0,0,0,0,0),\nonumber
\eeq
with $V_H>M_{GUT}$, which leads to the $SU(6)_\S \times SU(6)_H\rightarrow SU(6)_\S \times SU(5)_H$ breaking of the global symmetry, resulting in the breaking of the $SU(6)$ local symmetry down to $SU(5)$. This demonstrates that the theory of \cite{BD} recovers the construction of \cite{Inoue,Anselm} at low energies, without invoking the fine-tuning. Moreover, the above construction is even more general and can be used to address other problems as well, such as the fermion mass hierarchies. 

Furthermore, it must be pointed out that the natural reason for the above-mentioned splitting of the superpotential \eqref{w1} can be the discrete symmetry \cite{BD,BDM}-\cite{zura}, or an anomalous $U(1)$ symmetry \cite{DP}.

Hence, in this framework, the MSSM Higgs doublets are the Nambu-Goldstones (NG) of the accidental $SU(6)_\S \times U(6)_H$ global symmetry, which remain strictly massless in the exact SUSY limit. After supersymmetry gets broken, almost all of the components of the NG sector acquire the weak scale masses, with the exception of one combination of the scalar doublets. It remains massless at the tree-level, due to being the true pseudo-Nambu-Goldstone particle. However, the radiative corrections lift its mass, leading to the electroweak symmetry breaking \cite{BDM}.

Later, it was observed in \cite{trip} that the generalization of the GIFT mechanism to orthogonal groups generically gives raise to the extra color-triplet pNGs of interesting properties. Namely, even though they acquire weak scale masses, they are harmless from the point of view of proton stability because of the automatic $TeV/M_{GUT}$ suppression of their Yukawa couplings to quarks and leptons. Moreover, this leads to the lifetime of the lightest triplet to be in the range from $1\text{sec}$ to $1\text{day}$, making it very interesting phenomenologically. Its lifetime is long enough in order to appear stable in the detector, if produced at the accelerator. However, the construction of \cite{trip} is essentially the generalization of the model of \cite{Anselm} to $SO(10)$ gauge group and therefore suffers from the same lack of rationale for the enhanced global symmetry. In particular, the $SO(10)$ gauge invariant theory is arranged in such a way that merely the Higgs sector enjoys the $SO(12)$ global symmetry, which is impossible to achieve without fine-tuning.

There exists an alternative solution to the doublet-triplet splitting problem \cite{georgi}, which gives rise to the weak scale color-triplets, with the phenomenology similar to the ones from the previous paragraph. However, their origin is very different. Namely, their properties are caused by the special structure of the matter couplings and are not the results of the GIFT mechanism.

In this work we propose the mechanism which makes it possible to enlarge the Nambu-Goldstone spectrum of the non-fine-tuned GIFT scenario to include color-triplets, while retaining all of the nice features of $SU(6)$ unification. In particular, in order not to spoil the unification of gauge coupling constants, the extension of the NG sector should come in full representations of $SU(5)$ and the number of color-triplets should be at most six, for the theory to stay asymptotically free. From these it follows that adding $5$ and $\bar{5}$ to the NG spectrum is the most natural generalization of the GIFT mechanism. Moreover, the couplings of the triplet NGs to the light quarks and leptons should be extremely suppressed because of the proton stability.

The rest of the letter is dedicated to the realization of this program in the following order: In the second section we consider the general criteria necessary for the accommodation of naturally light triplet particles. In the third section, we construct an explicit model which achieves the above-mentioned objective naturally and we discuss its phenomenology. We conclude with a brief discussion of the issues, related to the extensions of the gauge group.

\section*{II. General Consideration}
In this section we discuss the modification of the $SU(6)$ model \eqref{w1}, necessary for accommodating the naturally light triplets \footnote{Without extending the gauge group.}. The fact that, in supersymmetric theories, $SU(6)\rightarrow SU(5)$ gives rise to the $5+\bar{5}+1$ Nambu-Goldstone superfields suggests that the enhancement of the global symmetry to $SU(6) \times SU(6) \times SU(6)$, by the introduction of one more sector of fundamental and anti-fundamental Higgses, puts us on the right track. 

Let us begin by assuming that it is possible to achieve the following splitting of the Higgs superpotential in a natural way
\beq
W=W(\S)+W(H_1,\bar{H}_1)+W(H_2,\bar{H}_2).
\label{w2}
\eeq
It is obvious that \eqref{w2} enjoys the `accidental' global $SU(6)_\S \times SU(6)_{H_1} \times SU(6)_{H_2}$ symmetry, corresponding to the independent transformations in each sector.

The theory is assumed to be arranged in a way that gives rise to the following breaking of the accidental global symmetry
\beq
&&SU(6)_{H_{1,2}}\quad \rightarrow SU(5)_{H_{1,2}},\nn \\
&&SU(6)_\S \qquad \rightarrow SU(4)\times SU(2)\times U(1),
\label{sym}
\eeq
which leads to the breaking of the $SU(6)$ gauge group down to the standard model one. The $SU(3)\times SU(2)\times U(1)$ decomposition of uneaten Nambu-Goldstones is
\beq
&&2\times\( 1,2,\frac{1}{2} \)_{\S+H_{1}+H_{2}}+2\times \( 1,2,-\frac{1}{2} \)_{\S+\bar{H}_{1}+\bar{H}_{2}}\nn \\
&&+\( 3,1,-\frac{1}{3} \)_{H_1+H_2}+\( \bar{3},1,\frac{1}{3} \)_{\bar{H}_1+\bar{H}_2}\nn \\
&&+(1,1,0)_{H_1+H_2},
\label{png}
\eeq
where the subscripts indicate the sector from which the Nambu-Goldstone multiplet came from. Since the difference between \eqref{png} and the Higgs content of the MSSM comes in full representations of $SU(5)$, the unification of the gauge coupling constants is unaltered.

Having discussed the Higgs sector, let us focus on the one containing the matter. For simplicity, we consider only one generation of light `fermions'. They can be accommodated in the chiral, anomaly-free super-multiplets $15+\bar{6}+\bar{6}'$.

In order to generate an appropriate mass for each of them, we need the Yukawa couplings. Moreover, the participation of the anti- and fundamental Higgs multiplets is essential; otherwise, it is impossible to give masses to all the fermions \footnote{The Yukawa couplings \eqref{bma7} must explicitly break the accidentally enhanced global symmetry otherwise the Nambu-Goldstone Higgses will decouple from the matter sector and the standard model fermions will remain strictly massless.}. But, as one can see from \eqref{png}, the triplet pseudo-Goldstone comes purely from the sectors of $H$'s. This means that in general the couplings of the following form
\beq
15\bar{H}\bar{6},\quad \frac{1}{M}15\S \bar{H}\bar{6},\quad \text{etc.},
\label{bma7}
\eeq
will lead to the unsuppressed couplings of the triplet NG to quarks and leptons. This, on the other hand, is unacceptable from the point of view of the proton stability. However, the problem will be resolved if by some mechanism the interactions like \eqref{bma7} involve combinations of $H$'s in which the light triplet does not reside.

In order to see how this could work, imagine that the Higgs superpotential is such that $\<H_1\>=\<H_2\>$ in the supersymmetric vacuum. Then, the triplet/anti-triplet living in the combination $H_1+H_2$ will be eaten up by gauge fields and become heavy, while the ones living in the orthogonal combination stay massless in SUSY limit. Hence, as long as the fermion superfields have Yukawa couplings only to the eaten-up combination of $H$'s, the light triplet is decoupled from the matter sector. These can be achieved by imposing the permutation symmetry $(H_1,\bar{H}_1) \leftrightarrow (H_2,\bar{H}_2)$. But, one immediately realizes that this will make the separation of two sectors of fundamentals from each other (and hence the enhancement of the accidental symmetry, to $SU(6)^3$) seem impossible. 

In the next section we construct  the model where this separation is accomplished using additional singlets. Although, the approximate accidental symmetry ends up being $SU(6)_\S \times SU(12)_{H_1+H_2}\times O(2)_{singlets}$ instead of $SU(6)^3$, the pseudo-Goldstone spectrum is the desired one, as we are about to show.

\section*{IV. $SU(6) \times \mathbb{Z}^{(1)}_2\times \mathbb{Z}^{(2)}_2 \times \mathbb{P}$ Model}

The model in which the triplets appear naturally light and decoupled from the standard matter sector is formulated as follows. The Higgs sector consists of the following chiral superfields\footnote{The numbers in parentheses indicate the representation of $SU(6)$.}: $\S(35)$, $H_{1,2}(6)$, $\bar{H}_{1,2}(\bar{6})$, $S_{1,2}(1)$ and $Y_{1,2}(1)$. We impose the following two $\mathbb{Z}_2$ symmetries
\beq
\label{z21}
&\mathbb{Z}^{(1)}_2:&\quad (H_1,\bar{H}_1)\rightarrow  -(H_1,\bar{H}_1),\quad S_1\rightarrow -S_1,\\
&\mathbb{Z}^{(2)}_2:&\quad (H_2,\bar{H}_2)\rightarrow  -(H_2,\bar{H}_2),\quad S_2\rightarrow -S_2,
\label{z22}
\eeq
supplemented with the permutation symmetry
\beq
&\mathbb{P}:&(H_1,\bar{H}_1)\longleftrightarrow(H_2,\bar{H}_2), \quad S_1\longleftrightarrow S_2.
\label{perm}
\eeq
The most general $SU(6) \times \mathbb{Z}^{(1)}_2\times \mathbb{Z}^{(2)}_2 \times \mathbb{P}$ invariant renormalizable superpotential\footnote{Here we do not discuss the mechanism for decoupling the adjoint sector, since there exists a plethora of possibilities for doing this, to what our construction is insensitive.} built out of the above listed superfields is given by 
\begin{widetext}
\beq
W=W(\S)
+(M_H+\alpha Y_1+\alpha ' Y_2)(H_1\bar{H}_1+H_2\bar{H}_2)
+(M_S+\beta Y_1+\beta ' Y_2)(S_1 ^2+S_2 ^2)\nn\\
+\Lambda^2 Y_1+\Lambda '^2 Y_2+\mu Y_1 ^2+\mu ' Y_2 ^2+\gamma Y_1^3+\gamma ' Y_2^3+\rho Y_1 Y_2
+\sigma Y_1^2 Y_2+\sigma' Y_1 Y_2^2.
\eeq
\end{widetext}
Here, all the dimension-full constants are assumed to be around the GUT scale.
It is easy to see that this superpotential enjoys the $SU(6)_\S \times U(12)_{H_1+H_2}\times O(2)_{S_1+S_2}$ accidental global symmetry. The theory possesses the ground state with vanishing $\mathcal{F}$ and $\mathcal{D}$-terms in which the VEVs of $\S$'s and $Y$'s are fixed to be
\beq
\< \S\>&=&V_\S \text{diag}(1,1,1,1,-2,-2),\nn \\
\<Y_{1,2}\>&=&V_{Y_{1,2}},
\eeq
while the $H$'s and $S$'s satisfy the relations that are the consequence of the imposed discreet symmetries \eqref{z21}, \eqref{z22} and \eqref{perm}
\beq
\<H_1\bar{H_1}\>+\<H_2\bar{H_2}\>&=&2V_H^2,\nn \\
\<S_1^2\>+\<S_2^2\>&=&2V_S^2.
\eeq
For purposes which will become clear shortly, we break this degeneracy by choosing the vacuum preserving the permutation symmetry \eqref{perm}
\beq
\<H_1\>=\<\bar{H}_1\>=\<H_2\>=\<\bar{H}_2\>&=&V_H,\nn \\
\<S_1\>=\<S_2\>&=&V_S.
\label{hvac}
\eeq
All the vacuum expectation values are determined by the parameters of the superpotential and are of the order of GUT scale. These lead to the following symmetry breaking pattern
\beq
\label{symS}
&&O(2)_{S_1+S_2}\quad \quad \rightarrow \O, \\
\label{symH}
&&U(12)_{H_1+H_2} \quad \rightarrow SU(11)\times U(1),\\
\label{symSig}
&&SU(6)_{\S}\qquad \quad \rightarrow SU(4)\times SU(2)\times U(1).
\eeq
As a result of \eqref{symH} the gauge group undergoes $SU(6)\rightarrow SU(5)$ breaking, broken further by VEVs of $\S$'s down to $SU(3)\times SU(2)\times U(1)$. For the radiative stability of $\text{sin}^2 \theta_W$, it is important that \eqref{symH} takes place before \eqref{symSig}. This can be easily arranged for by selecting the various parameters in a way to give $V_H \gg V_\S$.

Now, let us discuss in detail the fate of the Nambu-Goldstone superfields of the spontaneously broken accidental symmetry. The adjoint Higgs sector \eqref{symSig} gives NGs with the following $SU(3)\times SU(2)\times U(1)$ decomposition
\beq
(3,2)_{-2/3}+(\bar{3},2)_{2/3}+(1,2)_{1/2}+(1,2)_{-1/2}.
\label{goldsig}
\eeq
While in the sector of $H$'s and $S$'s, due to \eqref{symS} and \eqref{symH}, we get
\beq
2\times (3,1)_{-1/3}&+&2\times (\bar{3},1)_{1/3}+2\times (1,2)_{1/2}\nn\\
&+&2\times (1,2)_{-1/2}+4\times (1,1).
\eeq
However, because of the above mentioned breaking of the gauge symmetry, due to the supersymmetric Higgs mechanism, the following NG modes are eaten up by heavy gauge fields
\beq
(3,2)_{-2/3}&+&(\bar{3},2)_{2/3}+(3,1)_{-1/3}+(\bar{3},1)_{1/3}\nn\\
&+&(1,2)_{1/2}+(1,2)_{-1/2}+(1,1)_0,
\eeq
leaving us with the following massless constituents of the spectrum
\beq
(3,1)_{-1/3}+(\bar{3},1)_{1/3}+2\times (1,2)_{1/2}\nn\\
+2\times (1,2)_{-1/2}+singlets.
\label{goldstone}
\eeq
It is obvious that because of \eqref{hvac} and \eqref{goldsig} the light triplet and antitriplet from \eqref{goldstone} live entirely in the sector of $H$'s and are given by
\beq
T=\frac{T_{H_1}-T_{H_2}}{\sqrt{2}},\qquad \bar{T}=\frac{\bar{T}_{\bar{H}_1}-\bar{T}_{\bar{H}_2}}{\sqrt{2}}.
\label{triplet}
\eeq
The doublet NGs, on the other hand, reside in the combination of $\Sigma$'s and $H$'s.

Having discussed the Higgs sector in detail, let us move on to the consideration of its Yukawa couplings to the matter superfields. 
Again, we consider only one generation of `fermions' in the same anomaly-free supermultiplet as in the previous section, with the following $SU(5)$ decomposition
\beq
&15&=10+5=(q+u^c+e^c)_{10}+(D+L^c)_5, \nn \\
&\bar{6}&=\bar{5}+1=(d^c+l)_{\bar{5}}+n, \nn \\
&\bar{6}'&=\bar{5}'+1'=(D^c+L)_{\bar{5}'}+n'.
\label{decomp}
\eeq
Once the $H$'s develop a VEV, only the standard MSSM matter multiplets $10\subset 15$, $\bar{5}\subset \bar{6}$ should remain massless. The rest of the particles must acquire masses $\sim M_{GUT}$ and decouple from the light states.

The superpotential of the lowest dimension responsible for the fermion masses and respecting imposed discrete symmetries is given by
\\
\beq
W=&&\frac{1}{M}15(S_1\bar{H_1}+S_2\bar{H_2})\bar{6}'\nn\\
&&+\frac{\mathcal{C}}{M ^2}\(15H_1(\S H_1)15+15H_2(\S H_2)15\)\nn\\
&&+\frac{\mathcal{C'}}{M^2}\(15H_1(\S H_1)15+15H_2(\S H_2)15\)\nn\\
&&+\frac{\mathcal{S}_1}{M^2}15(\S(S_1\bar{H}_1+S_2\bar{H}_2))\bar{6}\nn \\
&&+\frac{\mathcal{S}_2}{M^2}15(\S (S_1\bar{H}_1+S_2\bar{H}_2))\bar{6},
\label{couplings}
\eeq 
where $M$ denotes the energy scale $> M_{GUT}$.
The first term makes $\bar{5}'_{\bar{6}'}$ and $5_{15}$ multiplets \footnote{The subscript indicates the $SU(6)$ multiplet, the particle came from, see \eqref{decomp}.} to have masses $\sim M_{GUT}$. The second and third lines of \eqref{couplings} are responsible for the Yukawa couplings of up-quarks with Higgses, while the fourth line is responsible for the down-quark Yukawa couplings. The last term, on the other hand, is relevant only for neutrino masses. The important observation is that because of \eqref{hvac}, the triplet-Goldstones \eqref{triplet} do not have Yukawa couplings to quarks and leptons in supersymmetric limit.

It must be noted that, even though we have included non-renormalizable interactions \eqref{couplings}, it is straightforward to make the theory renormalizable by integrating in the extra fermions of mass $M$, see \cite{BDSBH}.

The supersymmetry is assumed to be broken in the hidden sector and communicated to the observed one by minimal supergravity. This can be accounted for by the introduction of additional soft terms in the scalar potential \cite{BFS}, \footnote{The presence of the squared $\mathcal{F}$ and $\mathcal{D}$ -terms is assumed as well.}
\beq
V_{SB}=AmW_3+BmW_2+m^2\sum_k |\phi_k|^2+\text{h.c.}.
\label{soft}
\eeq
Here, $A$ and $B$ are soft SUSY breaking parameters, $m$ denotes the supersymmetry breaking scale, $\phi_k$ includes every scalar field in the theory and $W_{2,3}$ stands for the quadratic and cubic parts of the superpotential respectively. The general consequence of this breaking is that the pseudo-Goldstones of the approximate global symmetry acquire masses of order $m$.

Interestingly, at the GUT scale, most of the constituents of the pNG sector were massless only because of being related to the true pNG by supersymmetry. Hence, the moment SUSY gets broken, they are going to acquire the tree-level masses of the weak scale. However, the true pNG bosons (two scalar doublets and one scalar triplet, we discard the singlets) get their masses only through the radiative corrections.

\subsection{Triplet Phenomenology}
The model proposed in this work has an interesting accelerator signature because of the appearance of the light color triplets in the spectrum \cite{trip}.

The colored constituents ($T$-multiplet) of the pseudo-Goldstone sector are a Dirac fermion and two scalars with the $SU(3)\times U(1)$ quantum numbers of a $d$-quark. Moreover, in the case of gravity mediated SUSY breaking one of the scalar triplets, which we denote by $\tilde{T}$, will be the lightest state in $T$. But, it must be emphasized that, in general,  this result depends on the explicit SUSY breaking mechanism in action.

It is straightforward to verify that the effect of \eqref{soft}, besides raising the masses of the light modes to $m$, is the relative shift of the vacuum expectation values of the fundamental Higgs fields 
\beq
\< H_1 \>-\< H_2 \>\sim m.
\label{shift}
\eeq
Here the proportionality constant is determined by the soft breaking parameters $A$ and $B$. This, on the other hand, breaks the permutation symmetry \eqref{perm} responsible for the decoupling of the light triplet from the matter fields. As a result, $T$-multiplet develops the Yukawa couplings to the quark-lepton supermultiplets with $m/M_{GUT}$ suppression relative to the standard Yukawa couplings \cite{trip}. The couplings that are allowed by the unbroken gauge symmetry are given by $QTL$ and $u^c Td^c$. This makes its phenomenology extremely interesting. In particular, because of the above mentioned suppression of the Yukawa couplings, the lightest triplet $\tilde{T}$ has sufficiently long lifetime (from $1$sec to $1$day, depending on the flavor sensitivity of the couplings) for having an interesting accelerator signature \cite{trip}. Namely, once pair produced at the collider, because of the long lifetime it will hadronize by picking up a light quark from the vacuum; and depending on which quark it is going to pair with, the formation of two types of spin-1/2 hadrons is possible
\beq
T^0\equiv(\tilde{T}d),\qquad T^+\equiv(\tilde{T}u).
\eeq
In case of the production both of them will appear as stable hadrons for $1$sec$-1$day, unless the mass splitting is sufficiently large for the heavier $T$-hadron to $\beta$-decay to the lighter one. It is needless to mention that the Large Hadron Collider has enough power to test the existence of such exotic hadrons in the near future.
\section*{V. Outlook}
In the previous section we constructed a non-fine-tuned model with color-triplet pNGs. In order to decouple it from the light fermions, we made an essential use  of the permutation symmetry among the Higgs multiplets where the triplet pNG resides. It is obvious that in order to do so we need triplets to descend from the same representation, in our case $6$'s. Consequently, in order to prevent  contradiction with the experiment, in any modification of our model the adjoint sector should not give rise to the uneaten $(3,1)$ pNG. The only other possibility is to take all the Higgs sectors to be adjoint. As a result, the imposition of the permutation symmetry will be possible, but in this case there would be a couple of uneaten $(3,2)$'s, spoiling the asymptotic freedom. 

The above-mentioned requirements are highly constraining, in the context of the $SU(N>6)$ gauge groups. In particular, there are only two possible ways to generalize our construction.

The first possibility would be the absence of the triplet NGs in the adjoint sector. However, while being effective in the $SU(6)$ model, in the case of large gauge groups it forces the symmetry breaking pattern by the adjoint Higgs to be
\beq
SU(N)\rightarrow SU(N-2)\times SU(2)\times U(1).
\eeq
As a result, at low energies the extension of the MSSM Higgs sector will not come in full representations of $SU(5)$, spoiling the unification of the gauge couplings.

The second possibility is to allow the presence of the triplet NGs in the adjoint Higgs sector, but in this case they must be completely eaten-up by gauge field. In other words, it is not allowed for the eaten-up triplet to reside in a combination of adjoint and fundamental sectors because of the above-mentioned issues with the applicability of the permutation symmetry. This can be achieved only if one considers the trivial generalization of our $SU(6)$ model to larger groups. Namely, the symmetry breaking pattern in the adjoint sector should be
\beq
SU(N)\rightarrow SU(4)\times SU(2)\times U(1),
\label{adj}
\eeq
while the sectors of anti- and fundamental develop the VEVs in the direction of the unbroken $SU(4)$ of \eqref{adj}.
However, in order to realize \eqref{adj} one needs to enlarge the adjoint sector by introducing extra fields  in the representation different from the adjoin, e.g. $N$ and $\bar{N}$. Moreover, it is necessary to include couplings like $N\S \bar{N}$, otherwise the additional fields will form a separate sector and \eqref{adj} will not take place. But, it is a straightforward exercise \cite{BCL} to show that these kinds of terms generically require $\< N \>=0$, and the introduction of extra superfields of large content and vanishing VEVs becomes necessary. This, on the other hand, may accidentally enlarge the pseudo-Goldstone spectrum further, unless one does some sort of fine-tuning. Regardless fine-tuned or not, since the symmetry breaking pattern of this model is identical to the one of the previous section, this scenario implies $SU(6)$ unification as well; which by itself is embedded in the larger group. Interestingly, this scenario is indistinguishable from the case of just $SU(6)$, modulo the presence of extra singlets at the weak scale.

Finally, we have considered only one generation of fermions in this work. However, the generalization to three families is quite straightforward and will be presented elsewhere, along with the generation of the mass hierarchy among light fermions.

\section*{Achnowledgements}
I am grateful to Gia Dvali for useful discussions and support. This work was supported by James Arthur Graduate Fellowship at NYU.




\end{document}